# Dissipation and fluctuations in nanoelectromechanical systems based on carbon nanotubes

I V Lebedeva[1], A A Knizhnik[2,3], A M Popov[4], Yu E Lozovik[4] and B V Potapkin[2,3]

[1]Moscow Institute of Physics and Technology, 141701, Institutskii pereulok, 9, Dolgoprudny, Moscow Region, Russia

[2]Kintech Lab Ltd, 123182, Kurchatov Sq., 1, Moscow, Russia

[3]RRC "Kurchatov Institute", 123182, Kurchatov Sq., 1, Moscow, Russia

[4]Institute of Spectroscopy, 142190, Troitsk, Moscow Region, Russia

E-mail: lebedeva@kintech.ru

**Abstract.** Tribological characteristics of nanotube-based nanoelectromechanical systems (NEMS) exemplified by a gigahertz oscillator are studied. Various factors that influence the tribological properties of the nanotube-based NEMS are quantitatively analyzed with the use of molecular dynamics calculations of the quality factor (Q-factor) of the gigahertz oscillator. We demonstrate that commensurability of the nanotube walls can increase the dissipation rate, while the structure of the wall ends and the nanotube length do not influence the Q-factor. It is shown that the dissipation rate depends on the interwall distance and the way of fixation of the outer wall and is significant in the case of a poor fixation for the nanotubes with a large interwall distance. Defects are found to strongly decrease the Q-factor due to the excitation of low-frequency vibrational modes. No universal correlation between the static friction forces and the energy dissipation rate is established. We propose an explanation of the obtained results on the basis of the classical theory of vibrational-translational relaxation. Significant thermodynamics fluctuations are revealed in the gigahertz oscillator by molecular dynamics simulations and analyzed in the framework of the fluctuation-dissipation theorem. Possibility of designing the NEMS with a desirable Q-factor and their applications are discussed on the basis of the above results.

# 1. Introduction

Recently, a number of nanoelectromechanical systems (NEMS) that employ carbon nanotube walls as movable elements [1, 2] were proposed. These devices include rotational and plain nanobearings [3], nanogear [4], electromechanical nanoswitch [5], nanoactuator [6], Brownian motor [7], memory cell [8, 9], nanobolt–nanonut pair [10-12], and gigahertz oscillator [13, 14]. The crucial issue for the practical use of these devices is the ability to control the motion of the nanotube walls [15]. Several methods to control the relative sliding of the walls were suggested. For example, a nanotube functionalized with chemically adsorbed atoms or molecules may be operated by a non-uniform electric field [15] and a metallic movable wall may be driven by an inhomogeneous magnetic field [16, 17]. Furthermore, nanomotors based on the relative rotation [18-20], sliding [20] or screw-like motion [20] of carbon nanotubes walls were created. Thus, studying the characteristics of the NEMS based on the motion of carbon nanotube walls is an actual problem.

We consider the operation of NEMS exemplified by a gigahertz oscillator based on a double-walled nanotube (DWNT) [13, 14]. The gigahertz oscillator is one of the simplest NEMS. Thus, it is useful as a model system to study the general features of energy dissipation at the nanometer scale [16, 17, 21-32]. The energy dissipation rate in the DWNT oscillator is governed by (1) temperature, (2) geometrical parameters of the nanotube (oscillation amplitude, wall length, interwall distance) and (3) structural characteristics of the walls (commensurability between the inner and outer walls, structural defects, structure of the wall ends). The high dissipation rate related to the excitation of the low-frequency vibrational modes was observed when the initial telescopic extension of the movable wall exceeded a certain threshold value [21, 22]. Thus, the oscillation with an amplitude exceeding this threshold value should be avoided in the gigahertz oscillator operation. In the present study, we consider the initial telescopic extension below the threshold value.

It was shown that the interwall dynamic friction force in DWNTs strongly depends on temperature [22-24]. This indicates that a fluctuational thermal roughness of the nanotube walls makes a considerable contribution to the dynamic friction force. Therefore, one might expect that the influence of the wall structure on the dissipation rate is more pronounced at low temperature. Although the effects of the wall structure were addressed in a number of studies [24-31], the role of temperature in these effects was not clarified.

A weak dependence of the dissipation rate on the DWNT length was found at room temperature with the initial telescopic extension of the movable wall almost equal to the DWNT length [28, 29]. However, such a large oscillation amplitude may result in a high excitation of the low-frequency vibrational modes [21, 22]. Thus, the influence of the nanotube length on the dissipation rate in the case of a small oscillation amplitude, which presumably corresponds to a low dissipation rate, needs further consideration. An interwall distance seems to have some effect on friction in DWNTs. The static friction force in double-layer graphene systems [33] and DWNTs [34] substantially decreases with an increase in the interlayer/interwall distance. The lower dissipation rate was found for oscillators with the interwall distance close to 3.4 Å [16, 17, 22]. However, no detailed study of the influence of the interwall distance on the dissipation rate in DWNTs was performed.

A small difference in the dissipation rate was found between gigahertz oscillators with capped and open inner walls [26, 27]. However, in reality, open walls are generally terminated with chemically adsorbed molecules [35, 36]. Hydrogen functionalization of the walls was shown to slightly increase the dissipation rate at low temperature [27]. In addition, the experimentally obtained nanotubes contain structural defects, e.g., Stone–Wales defects [37] (with four adjacent hexagons switching to two pentagons and two heptagons) and vacancies [38]. It was revealed that defects may strongly modify the static friction force in double-layer graphite systems [33] and DWNTs [27]. Stone–Wales defects were shown to increase the energy dissipation in DWNTs at low temperature [27]. The dynamic friction force in DWNTs at room temperature was found to increase with the number of vacancies in the outer wall [24]. Defects were demonstrated to induce the instability of oscillators based on DWNTs with a large interwall distance [30]. On the other hand, in [31], a single defect in the outer wall was shown to prevent the interwall rotation and, thus, to reduce the dissipation rate at low temperature. Therefore, a systematic study of the influence of the defects and structure of the nanotube ends on the dissipation rate in the gigahertz oscillators at different temperatures is necessary.

The results on the influence of commensurability of the nanotube walls on the dissipation rate are rather controversial [25, 26, 28]. Guo *et al.* [25] and Rivera *et al.* [28] found that the energy dissipation rate in a commensurate DWNT is much higher than in an incommensurate DWNT, while Tangney *et al.* [26] observed equal dissipation rates in commensurate and incommensurate DWNTs. Guo *et al.* [25] performed low temperature simulation of the nanotubes of different interwall distances with a low initial telescopic extension of the inner wall. Tangney *et al.* [26] considered the nanotubes of the same interwall distance with a high initial telescopic extension of the inner wall at room temperature. Thus, there are a number of reasons that could lead to the discrepancy in the obtained results [25, 26]. First, the contribution of the fluctuational thermal roughness of the nanotube walls to the dissipation rate may considerably exceed the contribution of the nanotube wall structure at high temperature. Next, the low-frequency vibrational modes may be highly excited at a large initial telescopic extension of the inner wall [21, 22]. In this case, the dissipation rate is likely to be insensitive to the wall structure. Finally, the difference between the dissipation rates in [25] may be due to the inequality of the interwall distance of the two DWNTs. To clarify the effect of commensurability of the nanotube walls on the energy dissipation rate all these possibilities must be analyzed and commensurate and incommensurate DWNTs must be studied under the same conditions.

Thus, quite a number of studies were performed to examine the influence of various factors on the energy dissipation in the DWNT oscillator [16, 17, 21-32]. However, the qualitative results of these studies are difficult to compare, since quantitative estimations of the dissipation rate were not obtained. Moreover, an interpretation of the results is often complicated by a simultaneous action of several factors. Thus, a systematic analysis of the factors that influence the tribological properties of the nanotube-based NEMS would be useful. To perform this analysis a universal quantity which characterizes the energy dissipation is required. As such a quantity, we propose the oscillator quality factor (Q-factor). The gigahertz oscillator is characterized by a strong dependence of the frequency on the oscillation amplitude [13, 14]. Thus, the application of this NEMS with a constant frequency requires that the oscillation be sustained at a constant amplitude. There are two possible ways of the practical realization of the nanotube oscillator: 1) to choose the system parameters that provide a high value of the Q-factor and, therefore, a slow frequency drift, 2) to sustain the oscillation amplitude by an external force. For the second case, it was shown that this external force is inversely proportional to the Q-factor of the oscillator [15]. Hence, the Q-factor is well suited for the quantitative analysis of the various factors that influence the tribological properties and, consequently, the operation of the nanotube-based NEMS. We perform a direct molecular dynamics (MD) calculation of the Q-factor of the DWNT oscillators. The influence of temperature, the structural characteristics of the walls and the geometrical parameters of DWNTs on the energy dissipation in the nanotube-based NEMS is systematically examined in terms of the values of the Q-factor.

The results obtained here for the dependences of the Q-factor on the system parameters are explained on the basis of the classical theory of vibrational-translational relaxation [39-41]. Two limiting regimes of energy dissipation are revealed for the DWNT oscillator. In the first regime, populations of the nanotube vibrational levels are almost in thermal equilibrium. It is the case of a low dissipation rate. In the second dissipation regime, the populations of the vibrational levels are out of equilibrium, and a number of the vibrational modes are highly excited. The latter case is characterized by a high dissipation rate and should be avoided in the NEMS operation. The crossover between these two regimes is determined by the geometrical parameters of the oscillator.

We also study the possibility to predict the dynamic tribological behaviour of the nanotube-based NEMS by their static properties. However, we find that the static friction forces poorly correlate with the energy dissipation rate.

According to the fluctuation-dissipation theorem [42-44], the energy dissipation rate is associated with the magnitude of thermodynamic fluctuations. The principal feature of NEMS that is related to a small number of atoms in the system is that fluctuations in NEMS are significant. Thus, thermodynamic fluctuations may substantially influence the operation of NEMS. For example, it was shown that thermodynamic fluctuations restrict the minimum size of an electromechanical nanothermometer based on the interaction of the nanotube walls [45] and an electromechanical nanorelay based on the relative motion of the nanotube walls [9]. Thermodynamic fluctuations were also considered in the case of directional motion in Brownian motors [7]. Here we study thermodynamic fluctuations in the gigahertz oscillator with the use of molecular dynamic simulations and analyze the results in the framework of the fluctuation-dissipation theorem [42-44]. We show that the thermal noise in the gigahertz oscillator has a substantially different spectral density compared to that in the previously studied beam nanoresonator [46, 47] and harmonic oscillator [48] as the gigahertz oscillator has no fundamental frequency. The analysis of the thermal noise in the gigahertz oscillator performed here in the framework of the fluctuation-dissipation theorem confirms the results of the molecular dynamics study of the fluctuations in this NEMS. The restrictions imposed by the fluctuations on the nanotube-based NEMS are also considered.

The paper is organized in the following way. The oscillator configurations and simulation methods are described in section 2. In section 3, we calculate the static friction forces. The results of the molecular dynamic calculations of the Q-factor are given in section 4 and discussed in section 5. In section 6, thermodynamic fluctuations in the gigahertz oscillator are examined. Finally, we make some conclusive remarks.

## 2. Methodology

The analysis of the tribological properties of the DWNT oscillator was performed using empirical interatomic potentials. Van der Waals interaction between the inner and outer wall atoms was described by the Lennard–Jones 12–6 potential $U = 4\varepsilon ((\sigma/r)^{12} - (\sigma/r)^6)$ with the parameters $\varepsilon_{CC}$ = 3.73 meV, $\sigma_{CC}$ = 3.40 Å and $\varepsilon_{CH}$ = 0.65 meV, $\sigma_{CH}$ = 2.59 Å for carbon-carbon and carbon-hydrogen interaction, respectively, obtained from the AMBER database [49] for aromatic carbon and hydrogen bonded to aromatic carbon. The parameters provide a consistent description of the pairwise carbon-carbon and carbon-hydrogen interactions. The cut-off distance of the Lennard–Jones potential was taken to be 12 Å. The covalent carbon-carbon and carbon-hydrogen interactions inside the walls were described by the empirical Brenner potential [50], which was shown to correctly reproduce the vibrational spectra of defect-free as well as defect-containing carbon nanotubes [51].

We considered several DWNT oscillators of the length from 2.4 to 6.3 nm. The nanotube walls were equal in length [15]. Both ends of the outer wall were open and not functionalized, so that the telescopic extension of the inner wall was possible at both ends. Three types of the inner wall were considered. Both ends of the type I inner wall were open and not functionalized. The type II inner wall had one end capped and the other end open and not functionalized. The type III inner wall had one end capped and the other end open and terminated with hydrogen atoms.

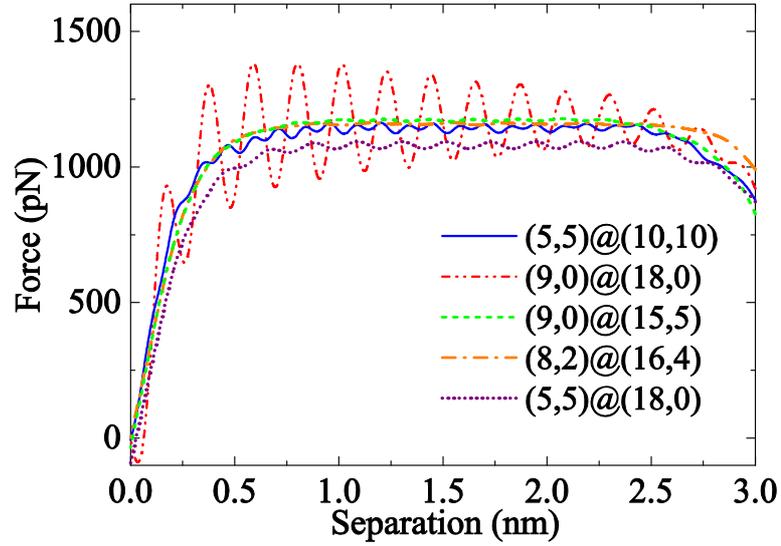

Figure 1. Calculated interwall van der Waals force of the (5,5)@(10,10) (blue solid line), (9,0)@(18,0) (red dash-dot-dotted line), (9,0)@(15,5) (green dashed line), (8,2)@(16,4) (orange dash-dotted line) and (5,5)@(18,0) (purple dotted line) DWNTs as a function of the center-of-mass separation of the walls. The DWNT length is 3.1 nm. The inner wall is of type I.

Figure 1 shows the total interwall van der Waals force as a function of the center-of-mass separation of the DWNT walls. Commensurate (5,5)@(10,10), (9,0)@(18,0) and (8,2)@(16,4) DWNTs and incommensurate (5,5)@(18,0) and (9,0)@(15,5) DWNTs were considered. The inner wall was of type I. The length of the walls was 3.1 nm. To obtain the dependences shown in figure 1 the nanotube walls were separately relaxed and then the inner wall was rigidly shifted along the common axis of the walls. The van der Waals force was found as the partial derivative of the interwall van der Waals energy with respect to the center-of-mass separation of the walls. The total van der Waals force of the interwall interaction in a DWNT can be broken down into two major components, a capillary force and a static friction force. The capillary force is defined as the average force that retracts back the wall which is telescopically pulled out of the DWNT to its original position. The capillary force approaches approximately 1200 pN for all the nanotubes under consideration. For large-diameter nanotubes, the interwall van der Waals energy should be proportional to the area of the overlap between the walls. Therefore, the capillary force is proportional to the nanotube diameter. A capillary force per unit nanotube diameter was found to be about 0.1 – 0.2 N/m in the experiments of Cumings *et. al.* [1] and Kis *et al.* [2]. We obtained the value of the capillary force per unit nanotube diameter of about 0.3 N/m. But one must take into account that the measurements [1, 2] were performed for large-diameter multi-walled nanotubes, whereas a non-linear dependence of the capillary force on the nanotube diameter should be expected for small-diameter nanotubes. The amplitude of the interwall van der Waals force oscillations, i.e. the static friction force, is higher in the commensurate DWNTs than that in the incommensurate DWNTs. The ratio of the static friction forces in the commensurate (5,5)@(10,10) and (9,0)@(18,0) DWNTs is about 1:14. This qualitatively agrees with the results of [25, 34, 52].

We performed microcanonical MD simulations of the operation of the DWNT oscillator. An in-house MD-kMC code was implemented. The code used the velocity Verlet algorithm and neighbour lists to improve the computing performance. The time step was 0.6 fs for the systems without hydrogen and 0.2 fs for the hydrogen-containing systems, which is about 100 times smaller than the period of thermal vibrations of carbon and hydrogen atoms, respectively. At the beginning of the MD simulations, the inner wall was pulled out along the DWNT axis by about 30% of its length and released with zero initial velocity. Generally, the outer wall was fixed at three atoms either in the wall body or at the wall edges. In some cases, in order to investigate the influence of the way of fixation on the dissipation rate, the outer wall was fixed at 10 atoms in the wall body. The relative fluctuations of the total energy of the system caused by numerical errors were less than 0.3% of the interwall van der Waals energy.

Our MD simulations of free oscillations demonstrated that the oscillations are damped by the interwall friction. The frequency of the damped oscillations of the (5,5)@(10,10) 2.4 – 6.3 nm nanotube-based oscillators is 43 – 105 GHz (see figure 2), in agreement with the other MD simulations [16, 17, 21–32].

We defined the Q-factor of the system as the ratio of the oscillation energy to the energy lost per one oscillation period. Since we performed the simulations of the system under adiabatic conditions, the energy dissipation rate was equal to the rate of the average thermal energy increase. The averaging of the thermal energy was performed over a time interval of 1 ps, as it must be much longer than the period of thermal vibrations (about 0.1 ps) and, on the other hand, much shorter than the period of the telescopic oscillation (about 10 ps). With the known dissipation rate, the Q-factor of the system was estimated.

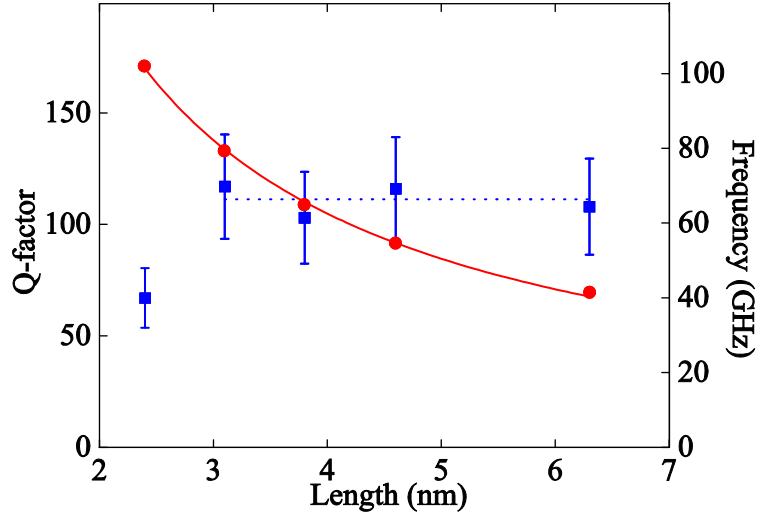

Figure 2. Calculated frequency (●) and Q-factor (■) of the (5,5)@(10,10) nanotube-based oscillator as functions of the nanotube length at a pre-heating temperature of 150 K. The inner wall is of type I. The outer wall is fixed at three atoms in the wall body. Solid line shows the best fit for the oscillation frequency as a function of the nanotube length. Dotted line is shown to guide the eye.

There are two contributions to the Q-factor calculation error. As mentioned above, the Q-factor depends on the oscillation amplitude. Thus, the oscillation damping leads to a change in the Q-factor. This effect becomes relevant at a long simulation time. On the other hand, there is also a stochastic contribution to the Q-factor calculation error related to thermodynamic fluctuations, which is high at a short simulation time and decreases with an increase in the simulation time. We took a simulation time of 500 ps, which minimizes the Q-factor calculation error in a single run and provides the accuracy of the Q-factor calculations within 20%. The temperature change over the simulation time was less than 9% at pre-heating temperatures of 50, 100, 150 and 300 K. At a pre-heating temperature of 0 K, the temperature increased to 1.5 – 15 K within the simulation time. The estimation performed in section 5 shows that the influence of quantum effects on the dissipation rate can be disregarded at temperature above 10 K.

## 3. Static force calculation

To analyze various contributions to the energy dissipation in the nanotube-based NEMS a universal quantity which characterizes the energy dissipation is required. We examined several static quantities which can characterize the dynamic tribological properties of the nanotube-based NEMS. In particular, a correlation between the dissipation rate and the static friction force is sometimes assumed [25, 27]. So we calculated the static friction force for a number of DWNTs.

The capillary force is zero for the wall of finite length sliding along the wall of infinite length. Thus, it is convenient to calculate the static friction force for such a system geometry. The inner wall length was taken to be 3.1 nm. The walls were separately relaxed and the inner wall was then rigidly shifted along the common axis of the walls. The interwall force was found as the partial derivative of the total interwall van der Waals energy with respect to the displacement of the inner wall. The static friction force was obtained as the maximum force that acts on the inner wall in the direction opposite to the direction of motion.

The influence of the nanotube wall structure on the static friction force was studied. The static friction force in DWNTs with non-chiral commensurate walls appeared to be much higher than that in DWNTs with incommensurate or chiral commensurate walls (see table 1). The type of the inner wall termination only slightly affects the static friction force (see table 2). The static friction force increases if structural defects are incorporated into the nanotube (see table 3).

**Table 1.** Calculated static friction force $F_s$, average absolute value of static force per atom $\langle |\vec{F}_\parallel| \rangle_{N,x}$ and Q-factor $Q$ of the nanotube-based oscillators with the type I inner wall at pre-heating temperatures of 0, 150 and 300 K. The DWNT length is 3.1 nm. The outer walls of the (5,5)@(10,10), (9,0)@(18,0) and (9,0)@(15,5) DWNTs are fixed at three atoms in the wall body. The outer walls of the (8,2)@(16,4) and (5,5)@(18,0) DWNTs are fixed at three atoms at the wall edges.

| DWNT | Interwall distance (Å) | $F_s$ (pN) | $\langle |\vec{F}_\parallel| \rangle_{N,x}$ (pN) | $Q$ (0 K) | $Q$ (150 K) | $Q$ (300 K) |
|---|---|---|---|---|---|---|
| (5,5)@(10,10) | 3.38 | 25 | 1.8 | 640 | 101 | 44 |
| (9,0)@(18,0) | 3.48 | 340 | 1.5 | 280 | 75 | 33 |
| (9,0)@(15,5) | 3.48 | 0.25 | 1.2 | 680 | 130 | 66 |
| (8,2)@(16,4) | 3.54 | 11 | 1.9 | 70 | 62 | 43 |
| (5,5)@(18,0) | 3.61 | 1.9 | 0.70 | 150 | 97 | 54 |

**Table 2.** Calculated static friction force $F_s$, average absolute value of static force per atom $\langle|\vec{F}_\||\rangle_{N,x}$ and Q-factor $Q$ of the (5,5)@(10,10) nanotube-based oscillator with the different type inner wall at pre-heating temperatures of 0, 150 and 300 K. The DWNT length is 3.1 nm. The outer wall is fixed at three atoms in the wall body.

| Nanotube type | $F_s$ (pN) | $\langle|\vec{F}_\||\rangle_{N,x}$ (pN) | $Q$ (0 K) | $Q$ (150 K) | $Q$ (300 K) |
|---|---|---|---|---|---|
| I. Both ends open and not functionalized | 25 | 1.8 | 640 | 101 | 44 |
| II. One end capped and the other end open and not functionalized | 28 | 1.6 | 520 | 100 | 48 |
| III. One end capped and the other end open and terminated with H atoms | 26 | 1.5 | 720 | 99 | 49 |

**Table 3.** Calculated static friction force $F_s$, average absolute value of static force per atom $\langle|\vec{F}_\||\rangle_{N,x}$ and Q-factor $Q$ of the (5,5)@(10,10) nanotube-based oscillator with the defect-containing type I inner wall at pre-heating temperatures of 0 and 150 K. The DWNT length is 3.1 nm. The outer wall is fixed at three atoms in the wall body.

| Defect | Location | $F_s$ (pN) | $\langle|\vec{F}_\||\rangle_{N,x}$ (pN) | $Q$ (0 K) | $Q$ (150 K) |
|---|---|---|---|---|---|
| no defects | | 25 | 1.8 | 640 | 101 |
| Stone–Wales defect | near the wall edge | 35 | 2.1 | 71 | 57 |
| Stone–Wales defect | in the middle of the wall | 36 | 2.0 | 89 | 66 |
| vacancy | near the wall edge | 36 | 1.9 | 68 | 53 |
| vacancy | in the middle of the wall | 40 | 2.6 | 57 | 57 |

For DWNTs with non-chiral (armchair/armchair, zigzag/zigzag, armchair/zigzag) walls, a good correlation between the static and dynamic friction force was found [25]. However, the static friction force in DWNTs with chiral commensurate walls is an order of magnitude lower than that in DWNTs with non-chiral commensurate walls [53] (see table 1). The point is that the unit cell in DWNTs with non-chiral commensurate walls is small and a large number of the movable-wall atoms are in equivalent positions with respect to the fixed wall. Forces acting on the unit cells are coherent and sum up to give a large total force. On the other hand, in DWNTs with incommensurate or chiral commensurate walls, only few atoms are in equivalent positions. This leads to a compensation of forces acting on different atoms in opposite directions. However, contributions of these forces to the dissipation rate are not compensated. So we suppose that the average absolute value of the static force (without the capillary component) acting on the movable-wall atom along the nanotube axis might be a better characteristic of the dissipation rate in DWNTs with incommensurate or chiral commensurate walls. The absolute value of the static force per atom should be averaged over all movable-wall atoms and axial displacements of the wall

$$\left\langle \left| \vec{F}_\| \right| \right\rangle_{N,x} = \frac{1}{N} \sum_{i=1}^{N} \left\langle \left| \vec{F}_{i,\|} \right| \right\rangle_x. \tag{1}$$

Here $\left\langle \left| \vec{F}_{i,\|} \right| \right\rangle_x$ is the force acting on the movable-wall atom $i$ along the nanotube axis averaged over axial displacements $x$ of the wall; $N$ is the number of the moveable-wall atoms.

As seen from table 1, the difference in the average absolute value of static force per atom between the DWNTs is not as high as the difference in the static friction force. Nevertheless, the average absolute value of static force per atom in the commensurate DWNTs is higher than that in the incommensurate DWNTs. The type of the inner wall termination almost does not influence the average absolute value of static force per atom (see table 2). Structural defects result in an increase of the average absolute value of static force per atom (see table 3). Below we compare the results obtained in the static force calculations with those obtained in the dynamic calculations of the Q-factor.

## 4. Dynamic Q-factor calculation

As discussed above, the Q-factor is an important characteristic of the gigahertz oscillator appropriate for the quantitative analysis of various factors that influence the tribological properties and, consequently, the operation of this NEMS. Hence, we performed the MD calculations of the Q-factor of the DWNT oscillator. The results of these calculations are presented in tables 1-4.

To compare our results with the others, we estimated the Q-factor for the systems considered in [22, 25, 26, 28, 29] from the time dependences of the oscillation amplitude given in those works. We assumed the interwall van der Waals energy of a long nanotube to be proportional to the center-of-mass separation of the nanotube walls. The found values of the Q-factor are given in table 5. As shown below, the data obtained in [22, 25, 26, 28, 29] agree reasonably well with our findings.

**Table 4.** Calculated Q-factor $Q$ of the nanotube-based oscillators with the type I inner wall and the outer wall fixed in different ways at pre-heating temperatures of 0, 150 and 300 K. The DWNT length is 3.1 nm.

| Way of fixation | (5,5)@(18,0) | | | (8,2)@(16,4) | | |
|---|---|---|---|---|---|---|
| | $Q$ (0 K) | $Q$ (150 K) | $Q$ (300 K) | $Q$ (0 K) | $Q$ (150 K) | $Q$ (300 K) |
| 3 atoms in the body | 120 | 56 | 54 | | 38 | 36 |
| 3 atoms at the edge | 150 | 97 | 54 | 70 | 62 | 43 |
| 10 atoms in the body and at the edge | 7200 | 970 | 140 | 108 | | |

The results of the Q-factor calculations for DWNT oscillators of different length with the initial telescopic extension of 30% of the nanotube length are presented in figure 2. A considerable decrease in the Q-factor is found for the DWNT that is shorter than 3 nm. For the longer DWNTs, the Q-factor is almost independent of the length. A weak dependence of the Q-factor on the DWNT length can be also found from the data obtained in [28, 29] with the initial telescopic extension of the inner wall almost equal to the nanotube length, at which a high excitation of the low-frequency vibrational modes [21, 22] might occur (see table 5). Thus, it seems that the DWNT length has no effect on the Q-factor, irrespective of the dissipation regime. The explanation of this fact is proposed in section 5.

The results presented in table 1 show that the Q-factor strongly depends on the interwall distance. Namely, at zero pre-heating temperature, the Q-factor of (5,5)@(10,10), (9,0)@(18,0) and (9,0)@(15,5) nanotube-based oscillators with the interwall distance close to 3.4 Å (which corresponds to the maximum absolute interwall van der Waals energy) is several times greater than the Q-factor of (5,5)@(18,0) and (8,2)@(16,4) nanotube-based oscillators with the interwall distance considerably above 3.4 Å. However, at room temperature, the Q-factor of the (5,5)@(18,0) and (8,2)@(16,4) nanotube-based oscillators becomes comparable to that of the (5,5)@(10,10), (9,0)@(18,0) and (9,0)@(15,5) nanotube-based oscillators. We suppose in section 5 that these results are associated with the fact that DWNTs with a relatively large interwall distance have a higher tendency for the excitation of the low-frequency vibrational modes.

**Table 5.** Oscillator Q-factors $Q$ calculated from the literature data. The values correspond to the time instance, at which the oscillation amplitude is 30% and 25% of the inner wall length for the data from [22, 26, 28, 29] and [25], respectively.

| Ref. | Nanotubes | Length (Å) | Initial extension (Å) | Interwall distance (Å) | Temperature (K) | $Q$ |
|---|---|---|---|---|---|---|
| [22] | (5,0)@(8,8) | 55/70 | 7.5 | 3.47 | 0 | 24 |
| [22] | (5,0)@(8,8) | 55/70 | 7.5 | 3.47 | 60 | 5.8 |
| [22] | (5,0)@(8,8) | 55/70 | 7.5 | 3.47 | 120 | 2.5 |
| [22] | (5,0)@(8,8) | 55/70 | 7.5 | 3.47 | 180 | 2.0 |
| [25] | (5,5)@(18,0) | 30 | 7.5 | 3.61 | 8 | 830 |
| [25] | (5,5)@(10,10) | 30 | 7.5 | 3.38 | 8 | 330 |
| [26] | (9,0)@(18,0) | 100 | 80 | 3.48 | 300 | 22 |
| [26] | (9,0)@(15,5) | 100 | 80 | 3.48 | 300 | 22 |
| [28, 29] | (7,0)@(9,9) | 122 | 119 | 3.36 | 300 | 7.6 |
| [28, 29] | (7,0)@(9,9) | 246 | 243 | 3.36 | 300 | 5.9 |
| [28, 29] | (7,0)@(9,9) | 369 | 366 | 3.36 | 300 | 6.5 |
| [28, 29] | (7,0)@(9,9) | 493 | 490 | 3.36 | 300 | 7.5 |

It also turned out that the dissipation rate in the gigahertz oscillator can considerably depend on how the outer wall is fixed (see table 4). In fact, in some cases of fixing three atoms in the outer wall body, a resonance between the telescopic oscillation of the (5,5)@(18,0) DWNT and some low-frequency nanotube vibrational modes gives rise to beats (see figure 3). Thus, these vibrational modes are highly excited. If, however, three other atoms are fixed (e.g., at the outer wall edges) the beats may be avoided and the dissipation rate decreases (see table 4). The Q-factor of the (5,5)@(18,0) nanotube-based oscillator can also be significantly increased by fixing a larger number of atoms (see table 4). The Q-factor of the (8,2)@(16,4) nanotube-based oscillator is less sensitive to the number of fixed atoms. We believe that an increase of the Q-factor of the nanotube-based NEMS may be achieved by the immobile wall fixation as a result of its interaction with an external medium. This possibility will be considered in a separate paper.

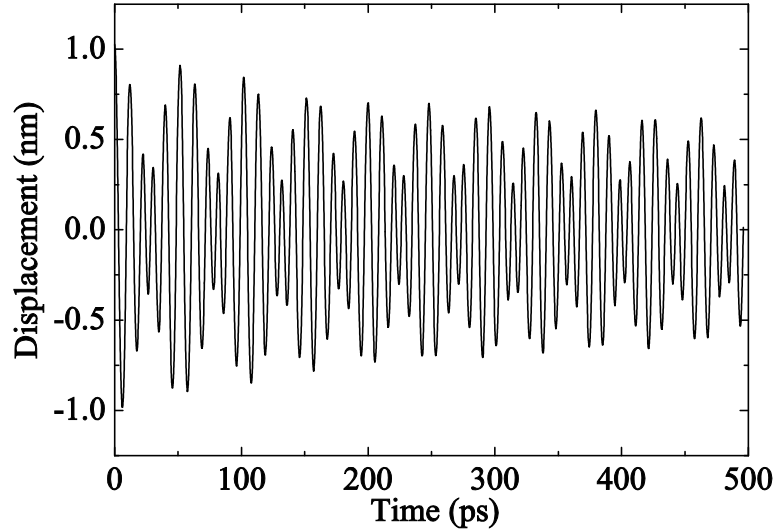

Figure 3. Calculated displacement of the movable wall of the (5,5)@(18,0) nanotube-based oscillator as a function of time at a pre-heating temperature of 150 K. The DWNT length is 3.1 nm. The inner wall is of type I. The outer wall is fixed at three atoms in the wall body.

To distinguish the effect of commensurability of the walls, we considered commensurate (9,0)@(18,0) and incommensurate (9,0)@(15,5) DWNTs of almost the same length, diameter and interwall distance. As seen from table 1, the Q-factor of the commensurate (9,0)@(18,0) DWNT with the initial telescopic extension of 30% of nanotube length is almost twice less than that of the incommensurate (9,0)@(15,5) DWNT, in agreement with [25, 28]. However, this result disagrees with Tangney *et al.* [26], who obtained equal dissipation rates for the (9,0)@(18,0) and (9,0)@(15,5) DWNTs at room temperature, yet with the initial telescopic extension of 80% of the nanotube length. We suppose that the discrepancy may be explained by a high excitation of the low-frequency vibrational modes [21, 22] that could occur in [26]. In fact, the Q-factor estimated from the data given by Tangney et *al.* [26] is considerably smaller than the value obtained in this work (see table 5). As discussed in section 5, the decrease in the Q-factor for commensurate DWNTs might be related to the interwall van der Waals force oscillations, which can lead to the resonant excitation of the vibrational modes. Figure 1 shows that the amplitude of the interwall van der Waals force oscillations in the (9,0)@(18,0) DWNT is considerably greater than that in the other considered DWNTs. This is the case for all (n,0)@(n+9,0) DWNTs compared to other DWNTs [52]. Therefore, the decrease in the Q-factor due to commensurability of the nanotube walls can be significant for all (n,0)@(n+9,0) nanotube-based oscillators. However, the pronounced influence of commensurability of the walls on the dissipation rate is not evident for other commensurate DWNTs (see the classification scheme of commensurate DWNTs in [52]). On the one hand, the Q-factor of the (8,2)@(16,4) nanotube-based oscillator is less than that of the (5,5)@(18,0) nanotube-based oscillator with the greater interwall distance. On the other hand, the high value of the Q-factor is observed for the oscillator based on the commensurate (5,5)@(10,10) DWNT (see table 1). Thus, the influence of commensurability of the walls on the tribological properties of the nanotube-based NEMS needs further investigation.

The Q-factor weakly depends on how the inner wall is terminated (see table 2). The changes in the Q-factor are within the calculation error for all the structures under consideration with the capped, open and not functionalized, open and hydrogen-functionalized ends, in reasonable agreement with [26, 27].

Since the experimentally obtained nanotubes contain structural defects, e.g., Stone–Wales defects [37] (with four adjacent hexagons switching to two pentagons and two heptagons) and vacancies [38], their tribological properties may considerably differ from the properties of defect-free nanotubes. We considered DWNTs with a single Stone–Wales or vacancy defect in the inner wall. The found values of the Q-factor are given in table 3. The Stone–Wales and vacancy defects significantly decrease the Q-factor as compared to the defect-free nanotube, in agreement with [24, 27, 30]. The defect type and position have a small effect on the energy dissipation rate. This indicates that the decrease in the Q-factor is not associated with the local defect structure. Fourier transforms of the center-of-mass displacements of the rings of the atoms in the middle of the inner wall in a direction perpendicular to the telescopic oscillation are shown in figure 4 for the defect-free and defect-containing (5,5)@(10,10) DWNTs (at the ground state, these rings lie in the planes perpendicular to the wall axis). As seen from the spectra shown in figure 4, the defects lead to the high excitation of the low-frequency vibrational modes, in agreement with [30]. The possible mechanisms of the influence of defects on the dissipation rate are discussed in section 5.

The Q-factor of the defect-free (5,5)@(10,10), (9,0)@(18,0) and (9,0)@(15,5) DWNTs with the interwall distance close to 3.4 Å strongly increases with a decrease in temperature (see figure 5), in agreement with [22-24]. Figure 5 shows that the Q-factor is almost inversely proportional to temperature (e.g., $Q^{-1} = 5.7 \cdot 10^{-5} T[K] + 1.3 \cdot 10^{-3}$ for the (5,5)@(10,10) DWNT with the type III inner wall), which corresponds to a linear dependence of the dynamic friction force on temperature. It should be noted that the Q-factor does not go to infinity at zero temperature. This is because of some energy exchange between the telescopic oscillation of the movable wall and the other degrees of freedom existing at zero temperature even in the classical limit, which leads to a non-zero energy dissipation.

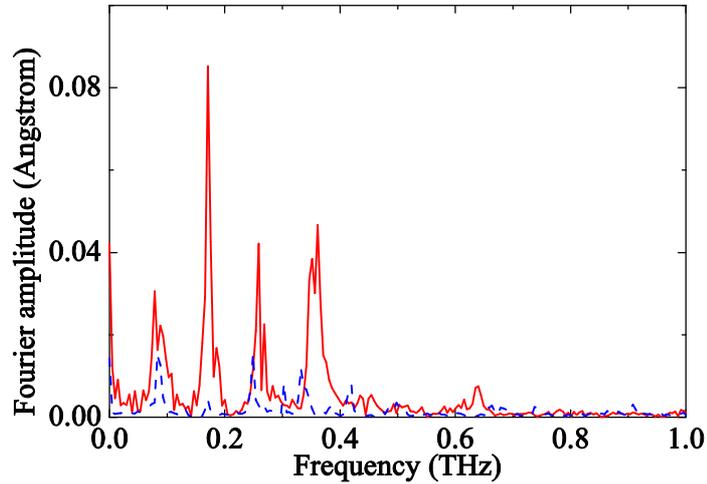

Figure 4. Calculated Fourier transforms of the center-of-mass displacements of the ring of the atoms in the middle of the inner wall in a direction perpendicular to the telescopic oscillation for the defect-free (blue dashed line) and defect-containing (a vacancy near the inner wall edge, red solid line) (5,5)@(10,10) nanotube-based oscillators at a pre-heating temperature of 150 K. The DWNT length is 3.1 nm. The defect-containing inner wall is of type I. The defect-free inner wall is of type III. The outer wall is fixed at three atoms in the wall body.

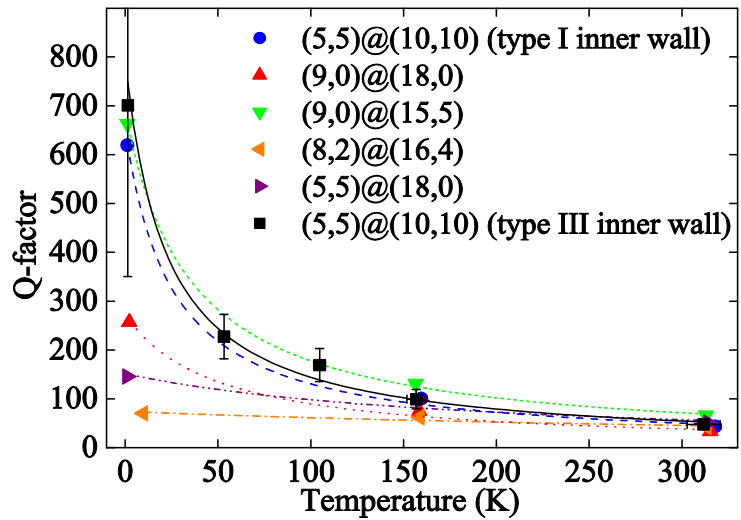

Figure 5. Calculated Q-factor of the (5,5)@(10,10) (●), (9,0)@(18,0) (▲), (9,0)@(15,5) (▼), (8,2)@(16,4) (◀), and (5,5)@(18,0) (▶) nanotube-based oscillators with the type I inner wall and of the (5,5)@(10,10) nanotube-based oscillator with the type III inner wall (■) as a function of temperature. The DWNT length is 3.1 nm. The outer walls of the (5,5)@(10,10), (9,0)@(18,0) and (9,0)@(15,5) DWNTs are fixed at three atoms in the wall body. The outer walls of the (8,2)@(16,4) and (5,5)@(18,0) DWNTs are fixed at three atoms at the wall edges. Solid line represents the best fit for the (5,5)@(10,10) nanotube-based oscillator with the type III inner wall. Other lines are shown to guide the eye.

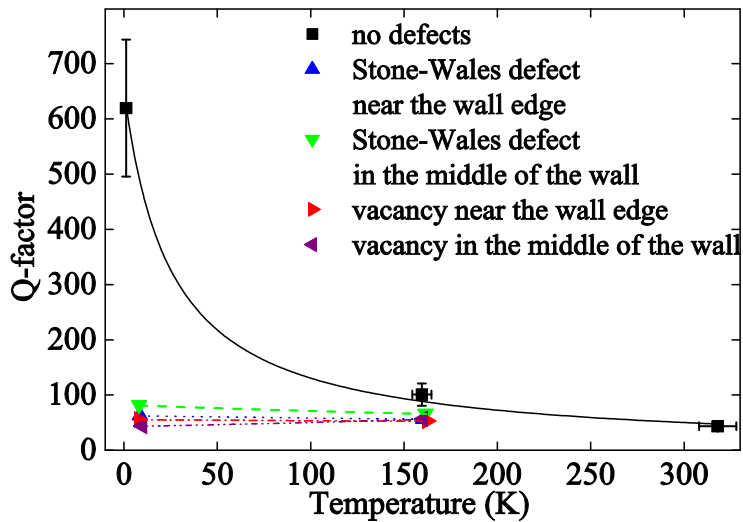

Figure 6. Calculated Q-factor of the defect-free (■) and defect-containing (5,5)@(10,10) nanotube-based oscillators as a function of temperature: a Stone–Wales defect near the wall edge (▲), a Stone–Wales defect in the middle of the inner wall (▼), a vacancy near the wall edge (▶), a vacancy in the middle of the inner wall (◀). The inner wall is of type I. The DWNT length is 3.1 nm. The outer wall is fixed at three atoms in the wall body. Lines are shown to guide the eye.

The strong temperature dependence of the Q-factor implies that there is a considerable contribution to friction due to a fluctuational thermal roughness of the walls. One might expect that the effect of the wall structure on the dissipation rate is suppressed at high temperature by thermal deformations of the walls. In this work, the effects of the structural characteristics of the nanotube walls were studied at pre-heating temperatures of 0, 150 and 300 K. As seen from table 1, the effect of commensurability of the nanotube walls is quite pronounced even at room temperature. The ratio of the Q-factor of the commensurate (9,0)@(18,0) and incommensurate (9,0)@(15,5) DWNTs is almost the same at high and low temperature. The type of the inner wall termination does not substantially affect the dissipation rate even at zero pre-heating temperature (see table 2).

The Q-factor of the (8,2)@(16,4) and (5,5)@(18,0) DWNTs with the interwall distance considerably above 3.4 Å and the defect-containing DWNTs weakly depends on temperature (see figures 5, 6). We suppose in section 5 that the weak temperature dependence of the Q-factor can be related to non-equilibrium populations of the vibrational modes in such DWNTs.

The static and dynamic calculations show a poor correlation between the static friction forces (total and average per one atom) and the Q-factor (see tables 1, 2, 3). The presence of defects leads to an increase of both the static and dynamic friction forces (see table 3). In the other considered cases of DWNT oscillators, the static friction forces almost do not correlate with the Q-factor. The observed absence of the correlation may be explained in the following way. As it is discussed in section 5, the energy dissipation rate is related to the resonant excitation of the nanotube vibrational modes. The static friction forces almost do not influence the resonance conditions and, hence, the dissipation rate.

## 5. Discussion of Q-factor dependences

To explain the dependencies of the Q-factor on temperature, the geometrical parameters of the nanotube and the structural characteristics of the walls we analyzed the kinetics of the energy transfer from the degree of freedom corresponding to the telescopic oscillation to the other vibrational modes of a DWNT. As shown later, the most significant contribution to the oscillation energy dissipation is provided by the excitation of the low-frequency vibrational modes. Among them, there are long-wave acoustic vibrational modes [54-56], squash modes [57] and non-translational relative vibrational modes of the walls [58]. A nanotube wall has four acoustic branches. These are the longitudinal (along the axis) branch, the degenerate transverse (perpendicular to the axis) branches and the twisting branch (along the nanotube circumference) [54]. The transverse acoustic branches are characterized the lowest sound velocity [54, 55].

We performed the MD simulations of the low-frequency vibrational modes of the (5,5)@(10,10) DWNT. At the beginning of the MD simulations, the nanotube atoms were given displacements corresponding to a particular vibrational mode. Fourier transforms of the center-of-mass displacements of the rings of the atoms in the middle of the inner wall in a direction perpendicular to the nanotube axis were used to determine the frequency of the excited vibrational mode. The lowest frequency of the transverse acoustic mode of the (5,5)@(10,10) DWNT of $L$ = 3 nm length was found to be 1.5 THz. The lowest frequency of the squash mode was shown to be about 1 THz. The obtained frequencies of the non-axial relative vibrations of the inner wall inside the outer wall are about 0.25 and 0.33 THz (the axial symmetry is violated by fixing of some outer wall atoms). The latter modes are strongly excited in the defect-containing oscillators (see figure 4). The obtained frequencies are in reasonable agreement with [54-58]. The high-frequency vibrational modes have the frequencies of 5 – 50 THz [54].

The energy transfer from the degree of freedom corresponding to the telescopic oscillation to the other nanotube vibrational modes can be described using the approach which is analogous to the classical theory of vibrational-translational relaxation [39-41]. According to this approach, the dissipation rate is related to the energy transfer from the translational motion of the movable wall to the nanotube vibrational modes

$$-\frac{dE}{dt} = -\sum_{\{n_i\},\{n_i'\}} \sum_i k_{\{n_i\},\{n_i'\}} N_{\{n_i\}} \Delta E_{\{n_i\},\{n_i'\}} \,. \tag{2}$$

Here $E$ is the oscillation energy; $i$ is the index of the nanotube vibrational mode; $n_i$ and $n_i'$ denote the quantum numbers corresponding the vibrational mode $i$; $k_{\{n_i\},\{n_i'\}}$ is the rate constant of the transition between the vibrational levels with the quantum numbers $\{n_i\}$ and $\{n_i'\}$; $N_{\{n_i\}}$ is the population of the vibrational level with the quantum numbers $\{n_i\}$; $\Delta E_{\{n_i\},\{n_i'\}}$ is the energy difference between the vibrational levels with the quantum numbers $\{n_i\}$ and $\{n_i'\}$.

In the harmonic approximation, the transitions occur only between neighbouring vibrational levels. Therefore, the dissipation rate is given by the equation

$$-\frac{dE}{dt} = \sum_{\{n_i\}} \sum_i \hbar\omega_i \left(k^i_{n_i,n_i+1} - k^i_{n_i,n_i-1}\right) N_{\{n_i\}} . \qquad (3)$$

Here $\hbar$ is Planck's constant; $\omega_i$ is the frequency of the vibrational mode $i$; $k^i_{n_i,n'_i}$ is the rate constant of the transition between the vibrational levels with the quantum numbers $n_i$ and $n'_i$. Moreover, the probability of the transition between the vibrational levels with the quantum numbers $n_i$ and $n'_i$ of the vibrational mode $i$ is proportional to the maximum of the quantum numbers $n_i$ and $n'_i$. Therefore,

$$k^i_{n_i,n_i+1} = k^i_{0,1}(n_i+1), \; k^i_{n_i,n_i-1} = k^i_{1,0} n_i . \qquad (4)$$

Finally, the expression for the dissipation rate takes the form

$$-\frac{dE}{dt} = \sum_{\{n_i\}} \sum_i \hbar\omega_i \left((n_i+1)k^i_{0,1} - n_i k^i_{1,0}\right) N_{\{n_i\}} . \qquad (5)$$

The rate constants $k^i_{0,1}$ and $k^i_{1,0}$ are independent of temperature but strongly depend on the Massey parameter $\omega_i \tau_c$, which is given by the product of the vibrational frequency $\omega_i$ and characteristic collision time $\tau_c$. If the Massey parameter is very large $\omega_i \tau_c \gg 1$ or very small $\omega_i \tau_c \ll 1$, the dissipation rate is negligible in any dissipation regime. The rate constants $k^i_{0,1}$ and $k^i_{1,0}$ sharply increase with approaching $\omega_i \tau_c \approx 1$, which is simply the resonance condition. The collision time is determined by the oscillation period, which is about 12 ps for the (5,5)@(10,10) DWNT of $L = 3$ nm length with the oscillation amplitude of 1 nm. As seen, the Massey parameter is close to 1 for the low-frequency vibrational modes and is very large ($\omega_i \tau_c \gg 1$) for the high-frequency vibrational modes. Thus, the dissipation is mainly provided by the low-frequency modes.

In thermal equilibrium at temperature $T$, the populations of the vibrational levels are given by

$$N_{\{n_i\}} = exp\left(-\sum_i \frac{n_i \hbar \omega_i}{k_B T}\right) \prod_i \left(1 - exp\left(-\frac{\hbar \omega_i}{k_B T}\right)\right), \qquad (6)$$

where $k_B$ is Boltzmann's constant.

So the dissipation rate is determined by

$$-\frac{dE}{dt} = \sum_i \hbar\omega_i \left( exp\left(\frac{\hbar\omega_i}{k_B T}\right) - 1 \right)^{-1} \left( k_{0,1}^i exp\left(\frac{\hbar\omega_i}{k_B T}\right) - k_{1,0}^i \right). \tag{7}$$

In the classical limit $\hbar\omega_i \ll k_B T$, one gets that the dissipation rate is linearly dependent on temperature

$$-\frac{dE}{dt} \approx \sum_i \left( k_B T \left( k_{0,1}^i - k_{1,0}^i \right) + \hbar\omega_i k_{1,0}^i \right) = AT + B, \tag{8}$$

where $A$ and $B$ are independent of temperature. Note that, in the case when the populations of the vibrational levels are in equilibrium at temperature corresponding to the classical limit, the considerable dependence of the dissipation rate on temperature is determined by the induced transitions (4). In the quantum limit $\hbar\omega_i \gg k_B T$, the dissipation rate does not depend on temperature

$$-\frac{dE}{dt} \approx \sum_i \hbar\omega_i k_{1,0}^i = B. \tag{9}$$

Equation (9) shows that, at low temperature, the dissipation rate is determined by the excitation to the first vibrational levels as a result of the spontaneous transitions.

For the low-frequency vibrational modes, which contribute most significantly to the dissipation rate, the condition $\hbar\omega_i < k_B T$ corresponds to temperature above 10 K. Therefore, if the Q-factor is almost inversely proportional to temperature, it implies that the populations of the vibrational levels are in thermal equilibrium (see figure 5). However, if the vibrational modes are out of thermal equilibrium, the populations of the vibrational levels are poorly related to temperature. As follows from equation (5), in this case, the dissipation rate weakly depends on temperature. This is observed for DWNTs containing defects or with the nanotube interwall distance above 3.4 Å (see figures 5, 6).

To explain the weak dependence of the Q-factor on the nanotube length we assume that the dissipation is mainly provided by a constant number of the long-wave acoustic vibrational modes and use the model potential with the constant interwall van der Waals force $F_W$, which is valid for a long nanotube [13, 14]. In our calculations, the ratio of the initial telescopic extension to the nanotube length was maintained. Therefore, the collision time is proportional to the nanotube length

$$\tau_c = 4\sqrt{\frac{2ma}{F_W}} \propto L. \tag{10}$$

Here $m \propto L$ is the mass of the movable wall; $a \propto L$ is the oscillation amplitude. The frequencies of the acoustic vibrational modes are inversely proportional to the nanotube length

$$f_n^{(k)} = \frac{nc_s^{(k)}}{2L}. \tag{11}$$

Here $k$ is the index of the acoustic branch; $n$ is the index of the vibrational mode; $c_s^{(k)}$ is the sound velocity of the acoustic branch $k$. So the Massey parameter of the acoustic vibrational modes is independent of the nanotube length. Under the assumption that the dissipation is mainly provided by a constant number of the long-wave acoustic vibrational modes, one gets that the dissipation rate (5) weakly depends on the nanotube length. The Q-factor is given by

$$Q = \frac{E}{\left|\frac{dE}{dt}\right|\tau_c}, \tag{12}$$

where $E \propto L$ is the oscillation energy. Thus, we obtain that the Q-factor of the gigahertz oscillator should not depend on the nanotube length.

Since the real potential of the gigahertz oscillator is complicated due to commensurability of the nanotube walls, edge regions and defects, $k_{0,1}^i$ and $k_{1,0}^i$ can be affected not only by the main characteristic collision time associated with the oscillation period but also by additional characteristic collision times. The additional collision times can be estimated as the times required for the movable wall to pass distances corresponding to the oscillations in the dependence of the static friction force on the displacement of the movable wall (see figure 1). Such distances may be determined by a size of a nanotube unit cell, an edge region or a defect and are 1 – 5 Å. The maximum velocity of the movable wall in our simulations is 6 Å/ps. Thus, one gets that the additional collision times in these cases are 0.1 – 1 ps. As seen, these collision times are resonant for the vibrational modes with frequencies of 1 – 10 THz. The influence of these collision times on the dissipation rate is determined by the magnitude of the potential non-uniformity. Therefore, the resonance between the low-frequency vibrational modes and the periodic non-uniformity of the interwall potential might provide the dependence of the Q-factor on commensurability of the nanotube walls. The weak effects of the structure of the wall ends and the nature of defects on the dissipation rate might be related to the similarity in the potential non-uniformity for all the structures under consideration.

Note that the influence of defects can be explained not only by the additional resonant collision time. In fact, the equilibrium position of the defect-containing inner wall is non-axial if the defect is inside the outer wall and axial if the defect-containing part of the inner wall is telescopically pulled out of the outer wall. Therefore, there is a force which acts on the defect-containing inner wall perpendicular to the axis and can excite non-translational vibrational modes of this wall.

## 6. Fluctuations in NEMS

According to the fluctuation-dissipation theorem [42-44], the energy dissipation rate is associated with the magnitude of thermodynamic fluctuations. We analyzed thermodynamic fluctuations in NEMS in the framework of the fluctuation-dissipation theorem. The first applicability condition of the fluctuation-dissipation theorem is that all degrees of freedom excluding one should be in thermal equilibrium. For the NEMS under consideration, this excluded degree of freedom is the telescopic oscillation of the movable wall of the gigahertz oscillator. Thus, the oscillator dynamics may be roughly described by the simplified one-dimensional Langevin equation of motion with allowance for the thermal noise

$$m\ddot{x}(t) + V'(x(t)) = F_{fr} + \xi(t) . \qquad (13)$$

Here $x$ is the displacement of the movable wall; $m$ is its mass; $V(x)$ is the interwall van der Waals potential; $F_{fr}$ is some friction force that models the energy dissipation; $\xi(t)$ is the thermal noise that represents random fluctuating forces. The left-hand side of equation (13) describes the deterministic part of the oscillator dynamics, while the right-hand side includes the effects of the thermal environment. The effects of energy dissipation and thermodynamic fluctuations are coupled by the common origin, the interaction of the oscillator with the microscopic degrees of freedom of the environment. The two terms in equation (13) that describe these effects become comparable when the energy loss over the time interval under consideration is close to a thermal quantum $k_B T$.

At a relatively low oscillation energy, the dynamic friction force is proportional to the oscillator velocity with the friction coefficient $\eta$, $F_{fr} = -\eta \dot{x}(t)$. Note that the linear dependence of the dynamic friction force on the velocity of the movable wall is the second applicability condition of the fluctuation-dissipation theorem.

In the approximation of the model potential with the constant interwall van der Waals force $F_W$, which is valid for a long nanotube [13, 14], the one-dimensional Langevin equation of motion takes form

$$m\ddot{x} + \eta\dot{x} + F_W sign(x) = \xi(t). \tag{14}$$

In the absence of the noise force ($\xi(t) \equiv 0$), the solution $x_0$ is the damping oscillation with the angular frequency

$$\Omega = \sqrt{\frac{\pi^2 F_W}{8ma}}, \tag{15}$$

where $a$ is the oscillation amplitude. Since the energy loss over the oscillation period $T_{osc}$ is given by

$$\Delta E = \int_0^{T_{osc}} \eta\dot{x}^2 dt = \frac{E}{Q} = \frac{F_W a}{Q}, \tag{16}$$

one also gets

$$\eta = \frac{3}{4\pi}\frac{m\Omega}{Q}. \tag{17}$$

Thermal noise $\xi(t)$ at temperature $T$ is a Gaussian white noise of zero mean $\langle\xi(t)\rangle = 0$, which satisfies the fluctuation-dissipation relation [42-44]

$$\langle\xi(t)\xi(t-\tau)\rangle = 2\eta k_B T \delta(\tau) = \frac{3}{2\pi}\frac{k_B T m\Omega}{Q}\delta(\tau). \tag{18}$$

Here $\delta(t)$ is the Dirac's delta-function. The spectral density of the noise force $\xi(t)$ in the gigahertz oscillator is given by

$$S_\xi = \frac{1}{2\pi}\int_{-\infty}^{+\infty}\langle\xi(t)\xi(t-\tau)\rangle exp(-i\omega\tau)dt = \frac{\eta k_B T}{\pi} = \frac{3k_B T m\Omega}{4\pi^2 Q}. \tag{19}$$

If the noise force is small enough, the displacement of the movable wall $x$ can be approximately represented as $x = x_0 + x_1$, where $x_1 \ll x_0$ corresponds to the noise-induced displacement of the movable wall

$$m\ddot{x}_1 + \eta\dot{x}_1 + F_W(sign(x_0 + x_1) - sign(x_0)) = \xi(t). \tag{20}$$

The analysis of the conditions where this approach is reasonable is presented below.

The force $F(x_0, x_1) = -F_W(sign(x_0 + x_1) - sign(x_0))$ is non-zero only when $x_0(t)$ crosses zero and leads to damping of $x_1$. So it can be treated as an effective friction force $-\eta'\dot{x}_1$ where $\eta'$ is determined by

$$\eta' \approx \frac{2m\Omega}{\pi}. \tag{21}$$

Since $\eta' \gg \eta$ (see equation (17)), the friction force $-\eta \dot{x}_1$ in equation (20) can be omitted

$$m\ddot{x}_1 + \eta' \dot{x}_1 = \xi(t). \tag{22}$$

Let us consider the harmonic external force $\xi(t) = \xi_\omega \exp(-i\omega t)$. Supposing $x_1(t) = x_\omega \exp(-i\omega t)$, one gets from equation (22) that

$$x_\omega = -\frac{\xi_\omega}{m\omega^2 \left(1 + i\frac{\eta'}{m\omega}\right)}. \tag{23}$$

Therefore, the spectral density of the amplitude noise is given by

$$S_x = \frac{\eta k_B T}{\pi m^2 \omega^4 \left(1 + (\eta'/m\omega)^2\right)}, \tag{24}$$

Note that this result directly follows from the fluctuation-dissipation theorem. It should also be mentioned that the obtained expression is quite different from those for the beam nanoresonator [46, 47] and harmonic oscillator [48], since the nanotube-based gigahertz oscillator has no fundamental frequency.

As follows from equation (24), the amplitude noise satisfies the relation

$$\begin{aligned}\left\langle (\Delta x_1)^2 \right\rangle &= 2\left\langle \left(x_1^2(0) - x_1(0)x_1(\tau)\right)\right\rangle = 2\int_{-\infty}^{+\infty} S_x(\omega)\left(1 - \exp(i\omega\tau)\right)d\omega \\ &= 2\frac{\eta k_B T}{\eta'^2}\left(|\tau| + \frac{m}{\eta'}\left(\exp\left(-\frac{\eta'}{m}|\tau|\right) - 1\right)\right)\end{aligned} \tag{25}$$

To compare the predictions of the fluctuation-dissipation theorem with the results of the MD simulations, we calculated the relative dispersion of the oscillation energy loss $\Delta E$. The average energy loss over a time interval of $\tau \ll T_{osc} Q$ is given by

$$\left\langle \Delta E \right\rangle = \frac{F_W a}{Q}\frac{\tau}{T_{osc}}. \tag{26}$$

The dispersion of the energy loss over the time interval of $\tau$ is determined by

$$\left\langle (\delta \Delta E)^2 \right\rangle = F_W^2 \left\langle (x_1(t) - x_1(t-\tau))^2 \right\rangle \approx 6\frac{k_B T F_W a}{Q}\left(\frac{\tau}{T_{osc}} + \frac{1}{4}\left(\exp\left(-4\frac{\tau}{T_{osc}}\right) - 1\right)\right). \tag{27}$$

Over the full simulation time $T_{sim}$,

$$\left\langle (\delta \Delta E)^2 \right\rangle \approx 6\frac{k_B T}{Q}\frac{T_{sim}}{T_{osc}} E \ll E^2, \tag{28}$$

which proves the condition $x_1 \ll x_0$ used above.

For $\tau = T_{osc}/2$, one gets

$$\delta = \frac{\sqrt{\langle (\delta \Delta E)^2 \rangle}}{\langle \Delta E \rangle} \approx 2.6\sqrt{\frac{k_B TQ}{E}} \ . \tag{29}$$

The analysis of the data obtained in the MD simulations implies the following relation for the relative dispersion of the energy loss $\Delta E$ over a half-period of the oscillation for defect-free DWNTs (see figure 7)

$$\delta \approx (2.46 \pm 0.06)\sqrt{\frac{k_B TQ}{E}} \ . \tag{30}$$

Moreover, the MD simulations conform the diffusion-like dependence $\langle (\delta \Delta E)^2 \rangle \propto \tau$ for $T_{osc}/2 < \tau \ll T_{osc}Q$. Thus, the fluctuation-dissipation theorem gives a reasonable estimate of the level of fluctuations in the system.

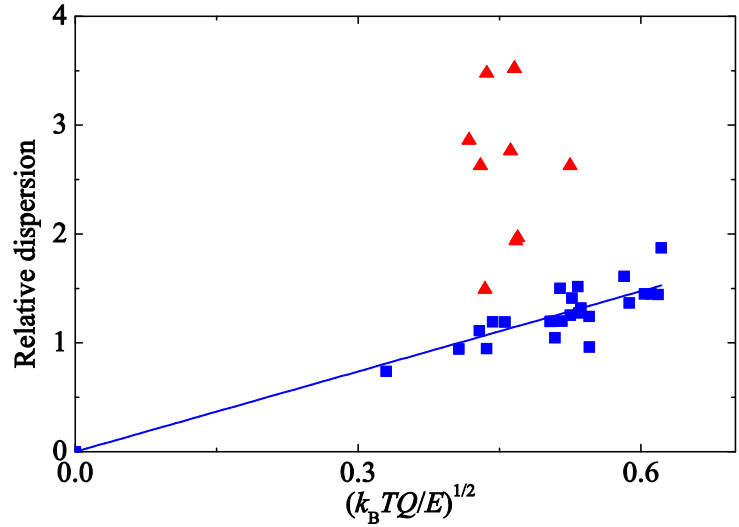

Figure 7. Calculated relative dispersion of energy loss $\Delta E$ over a half-period of the oscillation for the defect-free (■) (5,5)@(10,10), (9,0)@(18,0), (9,0)@(15,5) nanotube-based oscillators and the defect-containing (▲) (5,5)@(10,10) nanotube-based oscillators as a function of $(k_B TQ/E)^{1/2}$ at pre-heating temperatures of 50, 100, 150 and 300 K. Solid line shows the best fit for the defect-free oscillators.

As discussed in section 5, the nanotube vibrational modes are out of equilibrium in oscillators based on defect-containing DWNTs. Therefore, the second applicability condition of the fluctuation-dissipation theorem is not satisfied for such oscillators. In this case, our MD simulations show the substantial increase of the level of fluctuations in the system (see figure 7).

Since the oscillation energy loss per period in the system under consideration is comparable to the thermal quantum $k_BT$, significant fluctuations are observed (see figure 7). This is mainly associated with the small size of the system. The relative dispersion of the energy loss $\Delta E$ over a half-period for the defect-free 3.1 nm (5,5)@(10,10) nanotube-based oscillator is about 1.5 at a temperature of 300 K. With the same initial telescopic extension of the inner wall of 30% of the nanotube length, the relative dispersion of the energy loss $\Delta E$ over a half-period is below 0.5 for the oscillator that is longer than 30 nm. We suppose that the fluctuations can influence the possibility of the controllable operation mode of the gigahertz oscillator (this operation mode is considered in [15]). Thus, the fluctuations impose restrictions on the minimum size of the nanotube-based NEMS that may be used in applications.

## 7. Conclusions

We analyze the tribological characteristics of nanotube-based nanoelectromechanical systems (NEMS) by the example of the carbon nanotube-based gigahertz oscillator. The static and dynamic calculations show that, generally, the dynamic friction force does not correlate with the static friction forces.

The quantitative analysis of the factors that influence the tribological properties of the nanotube-based NEMS is performed through the molecular dynamic calculation of the Q-factor of the DWNT oscillator. The Q-factor proves to be insensitive to the structure of the wall ends. Moreover, the dissipation rate almost does not depend on the nanotube length if the ratio of the initial telescopic extension to the nanotube length is maintained. The dissipation rate is shown to increase with increasing the interwall distance at low temperature. It is found that commensurability of the walls of the (9,0)@(18,0) DWNT substantially decreases the Q-factor compared to that of the (9,0)@(15,5) DWNT with the same interwall distance. It is also demonstrated that the Q-factor depends on the way in which the outer wall is fixed. In some cases of fixing the outer wall at three atoms, beats can be observed between the telescopic oscillation of the (5,5)@(18,0) DWNT and the low-frequency nanotube vibrational modes. These beats can be eliminated by fixing the outer wall at other atoms. Stone–Wales and vacancy defects are shown to have a strong negative effect on the Q-factor. Fourier transforms of the center-of-mass displacements of the rings of the inner wall atoms in the direction perpendicular to the telescopic oscillation reveal that the defects lead to the high excitation of the low-frequency vibrational modes. Next, we find that the Q-factor of oscillators based on nanotubes containing defects or with the interwall distance above 3.4 Å depends on temperature considerably weaker than the Q-factor of oscillators based on defect-free nanotubes with the interwall distance of about 3.4 Å. In the latter case, the Q-factor is demonstrated to be inversely proportional to temperature. It is confirmed that the effect of the nanotube wall structure on the dissipation rate is pronounced even at room temperature.

Analysis of the kinetics of the energy transfer from the degree of freedom corresponding to the telescopic oscillation to the other nanotube vibrational modes is performed using the approach analogous to the classical theory of vibrational-translational relaxation [39-41]. The analysis shows that if the populations of the vibrational levels are in equilibrium, the dissipation rate should increase linearly with temperature. On the other hand, if the populations of the vibrational levels are out of equilibrium, the Q-factor should weakly depend on temperature. The weak dependence of the dissipation rate on the nanotube length is also explained. A qualitative explanation of the effects of defects and commensurability of the nanotube walls on the dissipation rate is suggested. This explanation is in agreement with the obtained vibrational spectra of defect-free and defect-containing nanotubes.

Moreover, the expression for the magnitude of fluctuations in the nanotube-based NEMS as a function of the Q-factor, the oscillation energy and temperature is found in the framework of the fluctuation-dissipation theorem [42-44]. We show that the spectral density of the thermal noise in the gigahertz oscillator is different from that in the previously studied beam nanoresonator [46, 47] and harmonic oscillator [48] as the gigahertz oscillator has no fundamental frequency. The molecular dynamic calculations reveal that the magnitude of the fluctuations conforms to the obtained expression for defect-free DWNT oscillators with the interwall distance of about 3.4 Å. This result implies the equilibrium populations of the vibrational levels for the defect-free DWNTs with the interwall distance of about 3.4 Å. For defect-containing DWNTs, the magnitude of the fluctuations is significantly higher than that predicted on the basis of the fluctuation-dissipation theorem, which implicitly confirms that the populations of the vibrational levels are out of equilibrium.

Based on the present investigation, some recommendations on the design of the nanotube-based NEMS with a desirable Q-factor can be made. In particular, slow damping is required for applications of the gigahertz oscillator. Thus, the device should operate at low temperature. A defect-free incommensurate DWNT with the interwall distance of about 3.4 Å should be used as the gigahertz oscillator. We believe that an increase of the Q-factor of the nanotube-based NEMS may be achieved by the immobile wall fixation as a result of its interaction with an external medium. On the other hand, a high energy dissipation is desirable for applications of a nanotube-based electromechanical memory cell [8, 9] to suppress oscillations of a movable wall after switching between the positions '0' and '1'. This implies high operation temperature and the use of a commensurate DWNT with a relatively large interwall distance. It is shown that thermodynamic fluctuations impose restrictions on the minimum size of the nanotube-based NEMS.

**Acknowledgments**

This work has been partially supported by the RFBR (AMP and YEL grants 08-02-00685 and 08-02-90049-Bel).